\documentclass[twocolumn, prx,reprint, amsmath,amssymb,showpacs,superscriptaddress,longbibliography]{revtex4-1}

\usepackage{graphicx}
\usepackage{dcolumn}
\usepackage{bm}
\usepackage{graphicx}
\usepackage{epsfig}
\usepackage{epsf}
\usepackage{amssymb}
\usepackage{amsmath}
\usepackage{amsthm}
\usepackage{multirow}
\usepackage{cases}
\usepackage[colorlinks=true,linkcolor=blue,citecolor=blue,pdfauthor={ },pdftitle={ },pdfsubject={ },pdfkeywords={ }]{hyperref}
\usepackage{pgfplots}
\usepackage{float}

\begin{document}

\title{Mesoscopic magnetic resonance spectroscopy with a remote spin sensor}

\author{Tianyu Xie}
\affiliation{CAS Key Laboratory of Microscale Magnetic Resonance and Department of Modern Physics, \\University of Science and Technology of China (USTC), Hefei 230026, China}
\affiliation{Hefei National Laboratory for Physical Sciences at the Microscale, USTC}

\author{Fazhan Shi}
\affiliation{CAS Key Laboratory of Microscale Magnetic Resonance and Department of Modern Physics, \\University of Science and Technology of China (USTC), Hefei 230026, China}
\affiliation{Hefei National Laboratory for Physical Sciences at the Microscale, USTC}
\affiliation{Synergetic Innovation Center of Quantum Information and Quantum Physics, USTC}

\author{Sanyou Chen}
\affiliation{CAS Key Laboratory of Microscale Magnetic Resonance and Department of Modern Physics, \\University of Science and Technology of China (USTC), Hefei 230026, China}
\affiliation{Synergetic Innovation Center of Quantum Information and Quantum Physics, USTC}

\author{Maosen Guo}
\affiliation{CAS Key Laboratory of Microscale Magnetic Resonance and Department of Modern Physics, \\University of Science and Technology of China (USTC), Hefei 230026, China}

\author{Yisheng Chen}
\affiliation{CAS Key Laboratory of Microscale Magnetic Resonance and Department of Modern Physics, \\University of Science and Technology of China (USTC), Hefei 230026, China}

\author{Yixing Zhang}
\affiliation{CAS Key Laboratory of Microscale Magnetic Resonance and Department of Modern Physics, \\University of Science and Technology of China (USTC), Hefei 230026, China}

\author{Yu Yang}
\affiliation{CAS Key Laboratory of Microscale Magnetic Resonance and Department of Modern Physics, \\University of Science and Technology of China (USTC), Hefei 230026, China}

\author{Xingyu Gao}
\affiliation{CAS Key Laboratory of Microscale Magnetic Resonance and Department of Modern Physics, \\University of Science and Technology of China (USTC), Hefei 230026, China}

\author{Xi Kong}
\affiliation{CAS Key Laboratory of Microscale Magnetic Resonance and Department of Modern Physics, \\University of Science and Technology of China (USTC), Hefei 230026, China}
\affiliation{Synergetic Innovation Center of Quantum Information and Quantum Physics, USTC}

\author{Pengfei Wang}
\affiliation{CAS Key Laboratory of Microscale Magnetic Resonance and Department of Modern Physics, \\University of Science and Technology of China (USTC), Hefei 230026, China}
\affiliation{Synergetic Innovation Center of Quantum Information and Quantum Physics, USTC}

\author{Kenichiro Tateishi}
\affiliation{RIKEN Nishina Center for Accelerator-Based Science, Wako, Saitama 351-0198, Japan}

\author{Tomohiro Uesaka}
\affiliation{RIKEN Nishina Center for Accelerator-Based Science, Wako, Saitama 351-0198, Japan}

\author{Ya Wang}
\affiliation{CAS Key Laboratory of Microscale Magnetic Resonance and Department of Modern Physics, \\University of Science and Technology of China (USTC), Hefei 230026, China}
\affiliation{Synergetic Innovation Center of Quantum Information and Quantum Physics, USTC}

\author{Bo Zhang}
\email{bz8810@ustc.edu.cn}
\affiliation{CAS Key Laboratory of Microscale Magnetic Resonance and Department of Modern Physics, \\University of Science and Technology of China (USTC), Hefei 230026, China}

\author{Jiangfeng Du}
\email{djf@ustc.edu.cn}
\affiliation{CAS Key Laboratory of Microscale Magnetic Resonance and Department of Modern Physics, \\University of Science and Technology of China (USTC), Hefei 230026, China}
\affiliation{Hefei National Laboratory for Physical Sciences at the Microscale, USTC}
\affiliation{Synergetic Innovation Center of Quantum Information and Quantum Physics, USTC}


\begin{abstract}
Quantum sensing based on nitrogen-vacancy (NV) centers in diamond has been developed as a powerful tool for microscopic magnetic resonance. However, the reported sensor-to-sample distance is limited within tens of nanometers because the signal of spin fluctuation decreases cubically with the increasing distance. Here we extend the sensing distance to tens of micrometers by detecting spin polarization rather than spin fluctuation. We detected the mesoscopic magnetic resonance spectra of polarized electrons of a pentacene-doped crystal, measured its two typical decay times and observed the optically enhanced spin polarization. This work paves the way for the NV-based mesoscopic magnetic resonance spectroscopy and imaging at ambient conditions. 

\end{abstract}

\maketitle

\section{INTRODUCTION}
As one of the most important techniques, magnetic resonance spectroscopy finds broad applications in chemistry, biology and material science.
Nanoscale magnetic resonance based on optical detection of electron spin resonance of nitrogen-vacancy (NV) centers in diamond has recently received broad attention in the context of quantum sensing. Magnetic resonance spectroscopy with nanoscale organic samples \cite{1,2, wrachtrup2017science} and single molecules \cite{3,4} have been realized.
Until now, the majority of nanoscale experiments measured a statistical fluctuation magnetization of spins which is much stronger than the mean thermal magnetization ($M_z \propto B_0 / T $) with a nano-detection volume under the ambient conditions with the magnetic field of several hundred gauss.
However, the fluctuation signal reduces dramatically with increased distance between the NV sensor and the sample.
For the mesoscale quantum sensing, e.g., cellular-sized magnetic resonance, the thermal polarization magnetization is stronger than the fluctuation.
Additionally, higher polarization can be achieved via hyperpolarization approaches such as optically induced polarization \cite{op}, dynamic nuclear polarization (DNP) \cite{dnp,dnp1,dnp2}, and quantum-rotor-induced polarization \cite{qrip,qrip1}. The polarization signal can be dominant once the spin polarization $P$ is reasonably high (normally, $P\sim 10^{-2}$ for electron spins and $\sim 10^{-4}$ for nuclear spins) even for the nanoscale sensing.

Here we report a long-range sensing method by detecting of spin polarization, so that mesoscale sensing based on NV center can be realised. This spin polarization removes the power law dependence on the separation distance between the target ensemble and the NV sensor. To demonstrate the method, we detect the mean magnetic field created by optically polarized electron spins within a pentacene-doped crystal. The optically induced polarization is improved a thousandfold compared to the calculated thermal polarization at $\sim$500 G. This results in three orders of magnitude signal enhancement. With this method, we can detect the magnetic resonance spectra and measure its two typical decay times of the pentacene molecules doped in a crystal with the size of a few tens of micrometers. The long-range sensing method paves the way for mesoscopic quantum sensing in chemistry, biology and material science at ambient conditions.
\section{PRINCIPLE}

\begin{figure}
\includegraphics[width=1\columnwidth]{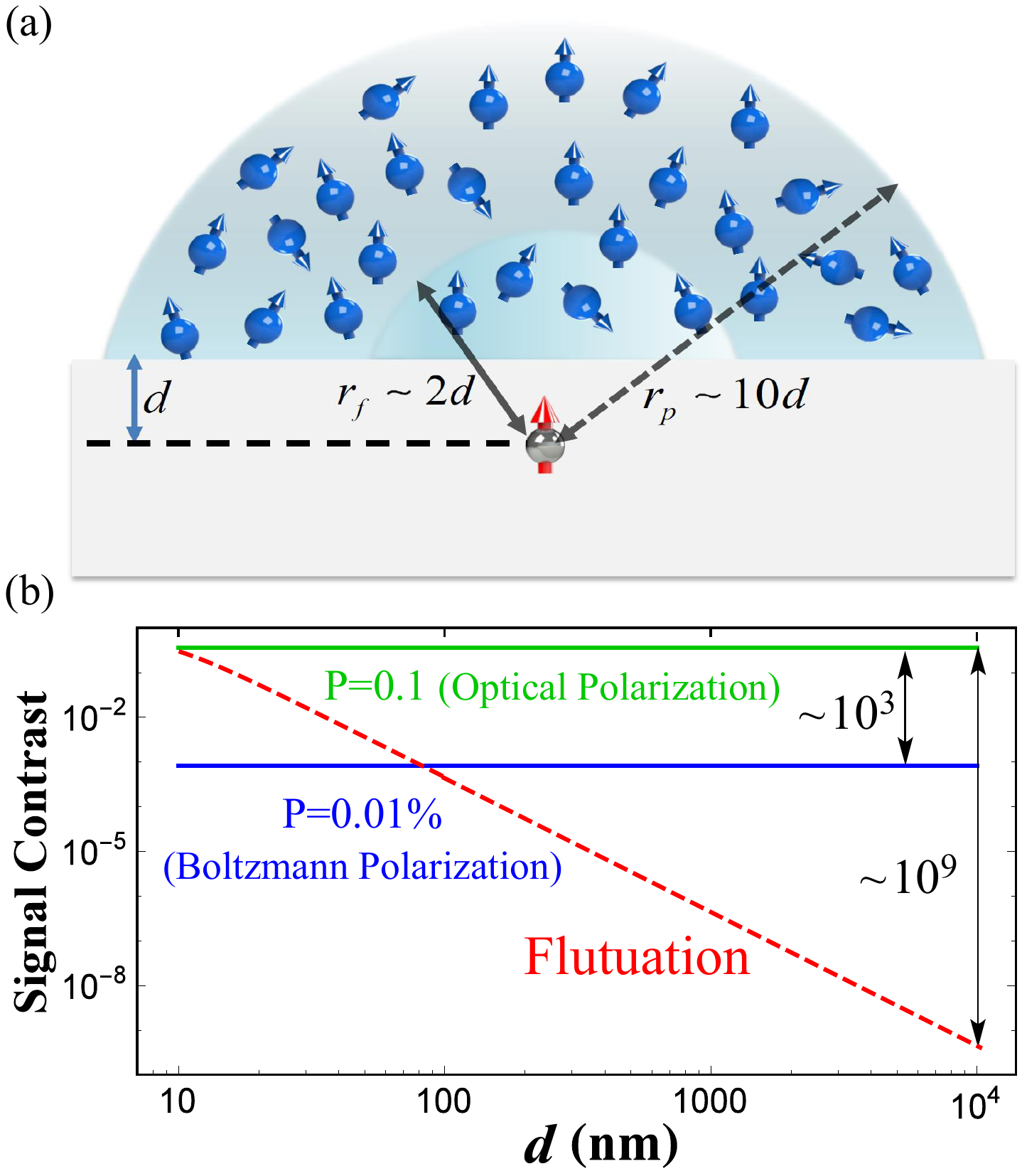}
\caption{Characterization of polarization signal (PS) and fluctuation signal (FS).
(a) A schematic illustration of detected spins within a large spherical sample shape ($r\rightarrow \infty$) probed by the NV sensor. The volume in which the spins generate 80$\%$ of the total signal is defined as the detection volume. For a single NV spin sensor with a depth $d$, the radius of PS detection volume, $r_{p}$, is about an order of magnitude larger than that of FS volume, $r_{f}$.
(b) The calculated signal contrasts (SC) of PS and FS as a function of NV-to-sample distance are illustrated. 
The signals with polarization of $P=10\%$ (green line) and $P=0.01\%$ (blue line) are the optically induced polarization and the Boltzmann polarization at $B_{0}$=500 G, respectively. The red dashed line represents the spin fluctuation signal. Here we make use of the parameters of pentacene spins for demonstration with accumulation time $t=7$ $\mu$s (the $m_{s}=+1$ relaxation time of pentacene molecule) and spin density $\rho \sim 10^{-3}$ nm$^{-3}$.  The NV center at a long distance from the sample is primarily sensitive to PS while the statistical spin fluctuation dominates the signal for the shallower NV center. The PS can be dominant for $P\geqslant 10^{-2}$ for the shallow NV.
}
    \label{range}
\end{figure}

A schematic of a non-interacting spin system probed via the NV sensor is shown in Fig. \ref{range}(a). The signal being probed originates from magnetic dipolar interaction between the NV center and the sample spins.

It can be viewed as a slowly varying magnetic field in the vicinity of the NV center. Whereas the fluctuation signal (FS) governed by its variance is in proportion to the square of the dipolar interaction (i.e. $\propto 1/r^{6}$), the polarization signal (PS) governed by its mean is linear to the dipolar interaction (i.e. $\propto 1/r^{3}$). Consequently, the calculated radius of PS detection volume, $r_{p}$, is about an order of magnitude larger than that of FS volume, $r_{f}$. After the integration of a spherical sample volume larger than detection one, the PS is independent of the NV depth while the FS decreases cubically with increased NV depth. By projecting the magnetic field onto the NV symmetry axis, we obtain the mean magnetic field $\bar{B}$ and the fluctuation one $ (\delta B)^{2} $ that NV can detect as shown below,
\begin{equation}
\bar{B} =  \dfrac{\mu_0 g_s}{6}\cdot M = \dfrac{4\pi c\rho P}{3} 
\label{1}
\end{equation}

\begin{equation}
\text{with}\:\:\: M = \mu_B\rho P, \;c \equiv \dfrac{\mu_0}{4\pi}\dfrac{g_s}{2}\mu_B
\label{2}
\end{equation}

\begin{equation}
(\delta B)^{2}=\dfrac{\pi c^{2}\rho (1-P^{2})}{4d^{3}}\doteq\dfrac{\pi c^{2}\rho}{4d^{3}}
\label{3}
\end{equation}
where $\mu_0$ is vacuum permeability, $\mu_B$ is Bohr magneton, $g_s$ is Land\'e  factor of electron spin, $\rho$ is the spin density of the sample, and $d$ is the NV depth below the diamond surface. $M$ denotes magnetization which equals magnetic moment per volume, i.e., $\mu_B\rho P$. $c$ is a constant and defined in Eqn. \ref{2} with the value 9.28 G$\cdot$nm$^{3}$ for free electron spin. 

Through interference measurement, $\bar{B}$ and $ (\delta B)^{2} $ can be recorded into the coherence of NV sensor. Then the coherence is transformed into the population of NV center which can be directly readout by fluorescence (See the detailed descriptions in \ref{sequence}). Therefore, the signal contrast (SC) is defined as the population difference of $m_{s}^{NV}=0$ between the measurements with/without the sample. In the limit of small signal contrast, it has the following relations with $\bar{B}$ and $ (\delta B)^{2} $.
\begin{equation}
SC_{pola} \approx 2\pi \gamma_e \bar{B} T/2
\label{4}
\end{equation}
\begin{equation}
SC_{fluc} \approx (2\pi \gamma_e \delta B T/2)^2
\label{5}
\end{equation}
$ \gamma_e $ is the gyromagnetic ratio of NV, i.e., 2.80 MHz/G, and $ T $ is the total time of interference measurement. The calculated signal contrast (SC) of PS and FS is mapped out as a function of NV depth $d$ in Fig. \ref{range}b.
\section{EXPERIMENTAL RESULTS AND DISCUSSION}
\subsection{Setup and pulse sequence description}
\label{sequence}
\begin{figure}
\includegraphics[width=1.0\columnwidth]{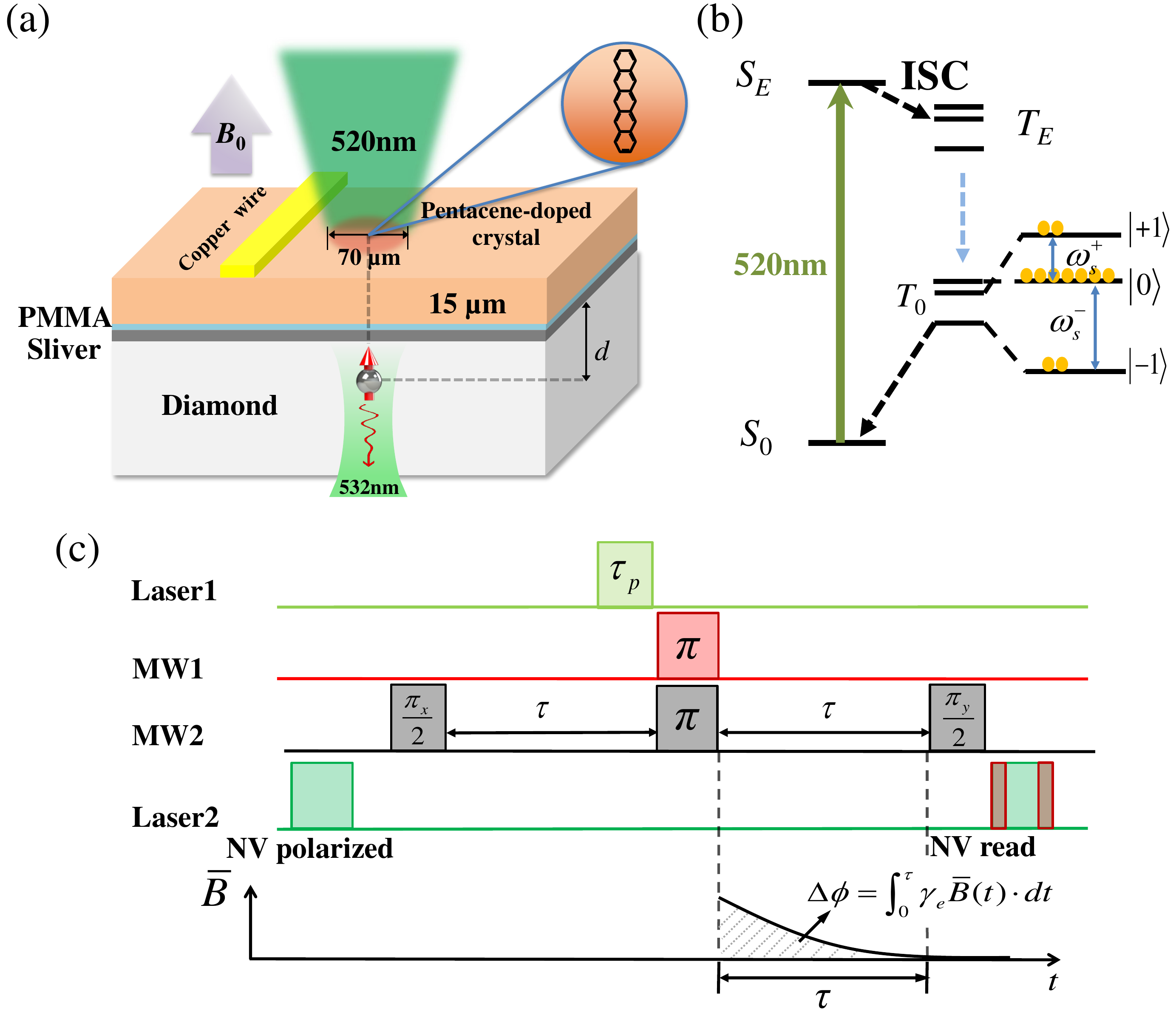}
\caption{(a) Schematic for optically induced polarization detection (OIPD) setup.
   (b) The electronic energy diagram of pentacene doped in a p-terphenyl crystal. Pentacene initially stays in the singlet ground state $S_{0}$ which is of magnetic-resonance silence. After being excited to the state $S_{E}$, it quickly decays into the $T_{E}$ triplet manifold through spin-selective intersystem crossing (ISC) with a typical time of $14.4$ ns \cite{5}. The resulting high spin polarization is preserved until it decays from $T_{0}$ into the singlet ground state $S_{0}$ in 10-100 $\mu s$ \cite{7} . The population in the state $|0\rangle$ of the triplet sublevels is much larger than the states $|\pm 1\rangle$ which can be infered from ref.\cite{6}.
   (c) The DEER pulse sequence for detecting the optically induced polarization. Laser 1 generates a 520-nm pulse to photo-excite the pentacene molecules, while MW1 produces a $\pi$ pulse for spins of pentacene molecules. Laser 2 creates 532-nm pulses to initialise NV centers and readout while the MW2 generates microwave pulses for NV spin echo.
   }
    \label{setup}
\end{figure}

In our experiment, we used the [111]-oriented NV center in a diamond chip, optically detected by a confocal microscope with a 532-nm laser excitation, shown in Fig. \ref{setup}(a). The NV sensor was a few microns below the diamond surface. The sample for detection was a single crystal of p-terphenyl doped with pentacene-$d_{14}$, 0.05 mol$\%$, where the long axis of the pentacene molecule was placed to align with the [111]-NV axis. The crystal thickness was 15 $\mu $m. Another 520nm-laser with the beam intensity of $\sim 10^{7}$ W/m$^{2}$ was applied to the crystal to optically induce the large population difference of two eigenstates which is defined as polarization, see the details below. Between the diamond and the crystal, there were a 150nm-silver layer and a 100nm-PMMA layer isolating the 532-nm and 520-nm laser beams as well as fluorescence generated from NV centers and the sample crystal. The copper wire on the top generated microwave fields to control both pentacene and NV.

The energy diagram is shown in Fig. \ref{2}(b). In our experiment, the sublevels of the triplet $T_0$ are not exactly the Zeeman eigenstates of external magnetic field $B_{0}$, but a weak mixing of them, i.e., $|m_{s}\rangle=\alpha|+1\rangle_{z}+\beta|0\rangle_{z}+\gamma|-1\rangle_{z}$, where $m_{s}=0$, $\pm 1$ and the subscript $z$ denotes the Zeeman eigenstates. See the detailed discussion in \ref{levels}. 

The optically induced polarization detection (OIPD) pulse sequence is shown in Fig. \ref{setup}(c) and comprises the following steps. (1) The NV spin state was initialized into $m_{NV}=0$ with a 1.5-$\mu s$ 532-nm laser pulse. (2) By introducing a $\pi/2$ pulse the NV spin was brought into the superposition state $(|0\rangle + |1\rangle)/\sqrt{2}$. It recorded $\bar{B}$ and $ (\delta B)^{2} $ generated by the sample during the following double-electron-electron resonance (DEER) pulse sequence. Synchronous with the $\pi$-pulse in the DEER sequence, a 520-nm laser pulse for time $\tau_{p}$ was applied to generate spin polarization of pentacene, defined as the population difference of sublevels $|0\rangle$ and $|+1\rangle$, $P=P_{0} - P_{+}$ (after the 520nm laser pulse, $P_{+} = P_{-}$). (3) After phase accumulation, a second $\pi/2$ pulse converted this phase into a measurable population that was read out by the final laser pulse. Note that owing to fast decay of the state $m_s=+1$ (7 $\mu$s in the Fig. \ref{5}), laser and
microwave pulses for pentacene spins can be applied before the first $\pi/2$ pulse for experimental
convenience. The pulse sequence was typically repeated for two million times to accumulate sufficient statistics, ensuring that the electron PS of pentacene was probed.

During the spin echo time $\tau$, the spin coherence of the NV spin revolves in the $x-y$ plane under the mean magnetic field $\bar{B}$ , and meanwhile shortens under the fluctuation one $ (\delta B)^{2} $. In the limit of small signal contrast, $\bar{B}$ is mainly stored in the imaginary part of the coherence $\int_{0}^{\tau} 2\pi\gamma_{e} \bar{B}(t)\cdot dt/2$ (PS) probed with two perpendicular $\pi/2$ pulses, while $ (\delta B)^{2} $ in the real part $(\int_{0}^{\tau}2\pi\gamma_{e} \delta B(t) \cdot dt/2)^{2}$ (FS) measured with two parallel $\pi/2$ pulses. However, FS is too small to be measured with deep NV ($\sim 3\times 10^{-10}$ for $d=10$ $\mu$m shown in Fig.\ref{range}(b)).
\subsection{Magnetic resonance spectroscopy of optically polarised electron spins}
\begin{figure}\centering
\includegraphics[width=1.0\columnwidth]{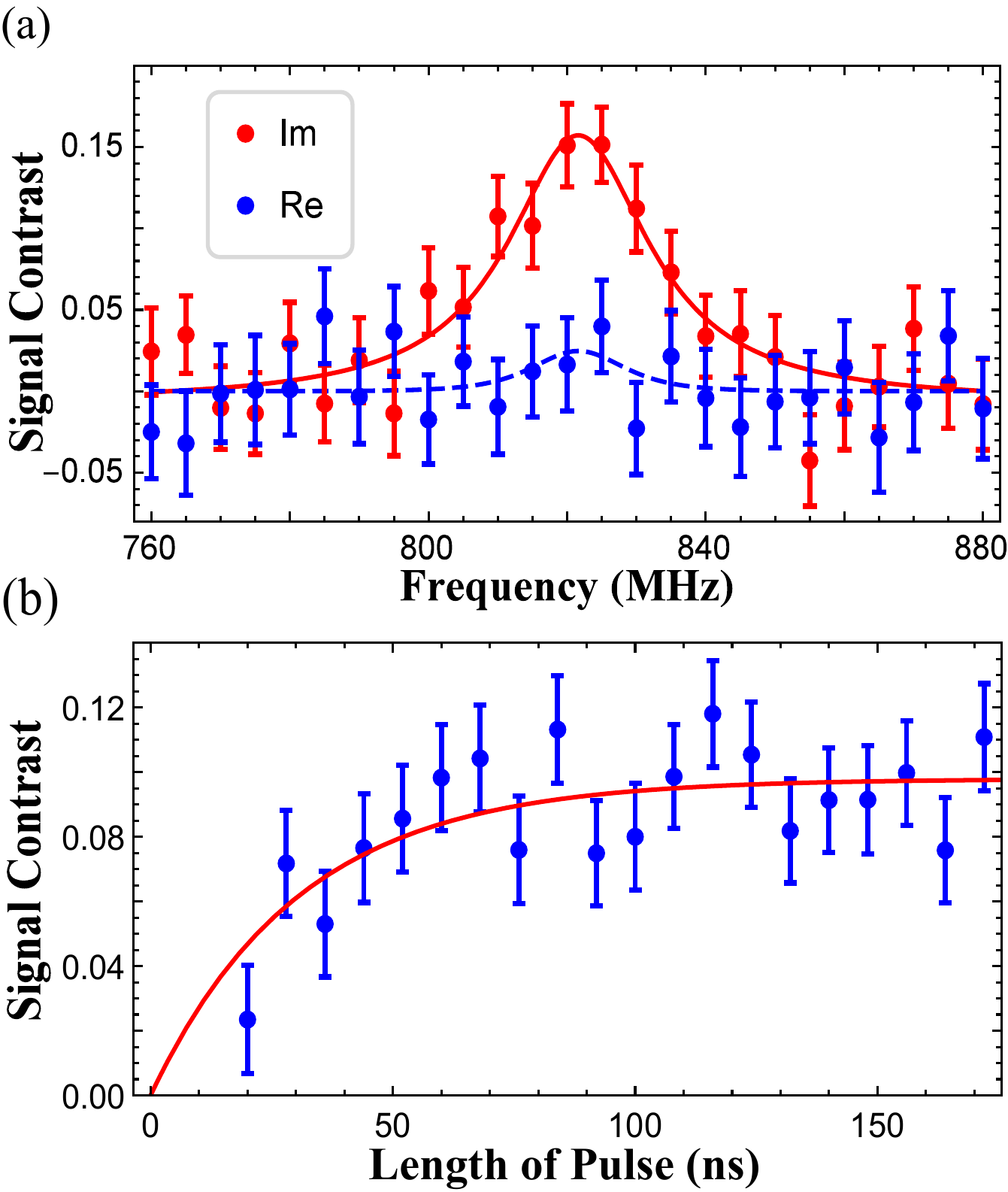}
\caption{(a) The ESR spectrum of pentacene is obtained through the OIPD pulse sequence. Red circle line (blue circle line) indicates the result of $(\pi/2)_{y(x)}$ pulse. The red solid line is the Lorentzian curvefit of the spectrum, from which the blue dashed line is calculated according to the fitting results.
   (b) The transition between the states $m_s=0$ and $m_s=+1$ by varying the length of MW1. 
   }
    \label{spectrum}
\end{figure}

We first recorded an electron spin resonance spectrum of pentacene with an NV depth of $d=12$ $\mu$m at the field $B_{0}=$512 G. The OIPD sequence was repeated with a sequential scan of $\omega_{MW1}$, in which the laser irradiation time $\tau_{p}=$ 1.5 $\mu$s, spin echo time $\tau=$ 21.6 $\mu$s. Fig. \ref{spectrum}(a) presents the results obtained from applying two different phases of the second $\pi/2$ pulse. As explained in the \ref{sequence}, for $(\pi/2)_{y}$ pulse, a strong peak (red circle) at $822\pm 1$ MHz is observed while the peak is absent for $(\pi/2)_{x}$ pulse.
 
As shown in Fig. \ref{spectrum}(b), the transition between the sublevels of $m_{s}=0$ and +1 of pentacene molecule was measured but no Rabi oscillation was observed. It indicates that the microwave field MW1 manipulating electron spins of pentacene is rather inhomogeneous. Besides, it can be inferred from Fig. \ref{spectrum}(b) that the maximum mixed state of $|0\rangle$ and $|+1\rangle$ is generated on the ensemble average after the MW1 irradiation of 80 ns in Step (2), and thus halves the magnetic field $\bar{B}$.
\subsection{Energy levels of triplet states versus magnetic field}
\label{levels}

\begin{figure}
\includegraphics[width=1\columnwidth]{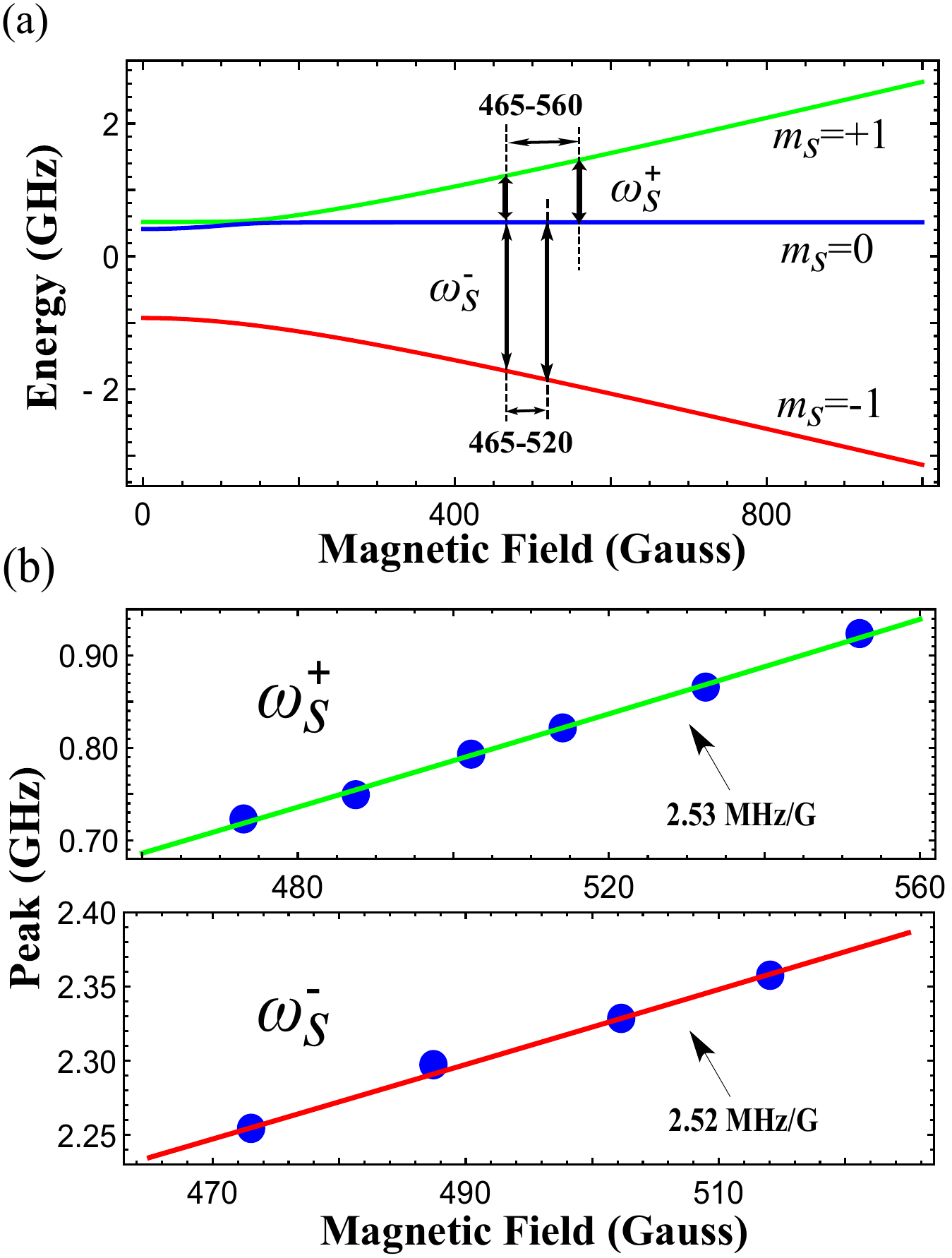}
\caption{(a) Energy level diagram of pentacene electron spin. Green, blue, and red solid lines represent the field-dependence of the spin energy sublevels, $m_S$ = 0 and $\pm$1, (choose $\theta$ = 8$^\text{o}$ and $\phi$ = 20$^\text{o}$). The data collected are in the interval between 465 and 560 Gauss for $\omega^+_S$, the transition between $m_S$ = +1 and 0, and in the interval between 465 and 520 Gauss for $\omega^-_S$, the one between $m_S$ = -1 and 0. (b) Transition frequencies $\omega^+_S$ and $\omega^-_S$ as a function of the magnetic field. Blue dots are experimental data, while solid lines are the calculated results of the corresponding transitions in (a). 
}
\label{gr}
\end{figure}

The Hamiltonian of triplet states of pentacene molecule under an external magnetic field is given by
\begin{equation}
\label{Eq:Hamiltonian}
\begin{aligned}
H = D S_z^2 + E(S_x^2 &- S_y^2)+\gamma_eB_0(\sin(\theta)\cos(\phi)S_x\\&+\sin(\theta)\sin(\phi)S_y+\cos(\theta)S_z)
\end{aligned}
\end{equation}
where $D$ and $E$ are the zero-field splitting (ZFS) parameters, $\gamma_e$ is the gyromagnetic ratio of the spin, and $B_0$ is the amplitude of the magnetic field. Here we designate the molecular axes of pentacene as follows: the out-of-plane axis as the x axis, the short in-plane axis as the y axis, and the long in-plane axis as the z axis. $\theta$ and $\phi$ are the spherical coordinates of the magnetic field in the frame. A small misalignment 
($\theta \thicksim 8^\text{o}$) results from the lattice structure ($\thicksim 2^\text{o}$) \cite{Lat} and the slight tilting angle caused by cutting ($\thicksim 5-10^\text{o}$), and the latter dominates. According to ref. \cite{8}, $D$ = -776.55 MHz and $E$ = -669.75 MHz (Note: the coordinate system in the ref. \cite{8} is different from the one used here, and we need perform a $\pi/2$ rotation about y axis to obtain the parameters here). Due to the transverse component $E$ of the ZFS (non-commute with the Zeeman interaction), it associates with spin mixing of Zeeman states, $|0\rangle_{z}$ and $|\pm1\rangle_{z}$, the energy levels are curved with respect to the magnetic field, shown in Fig. \ref{gr}(a). Consequently, instead of the gyromagnetic ratio of free electron, 2.80 MHz/G, the slopes of $\omega^+_S$ and $\omega^-_S$ are 2.53 MHz/G and 2.52 MHz/G in the vicinity of 500 G, respectively. As shown in Fig. \ref{gr}(b), the experimental data agrees well with the calculated results, indicating that the signal originates from the pentacene molecule and not from electron contamination on the surface of or within the diamond. 

\subsection{Dynamical study of triplet states in pentacene molecules}
\begin{figure}
\includegraphics[width=1\columnwidth]{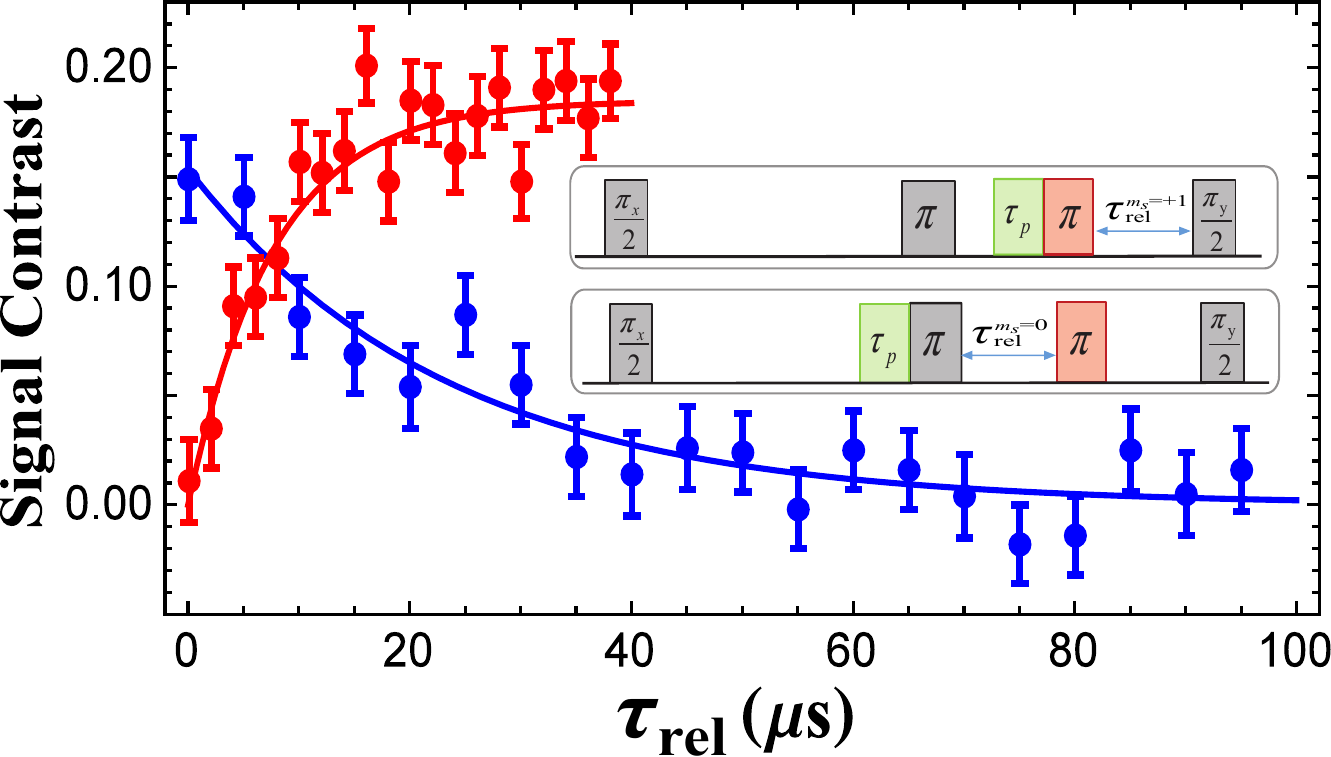}
\caption{Relaxation curves for $m_{s}=+1$ state (red dot) and $m_{s}=0$ state (blue dot). Inset: OIPD pulse sequences including relaxation intervals, $\tau_{rel}^{m_{s}=+1}$ and $\tau_{rel}^{m_{s}=0}$.}
\label{dyn}
\end{figure}

The dynamical property, such as the decay times of the pentacene $T_{0}$ manifold $m_{s}=+1$ and $m_{s}=0$, has been investigated in an adaptation to the OIPD sequence. A relaxation interval $\tau_{rel}^{m_{s}=+1}$ was inserted between the microwave irradiation for pentacene (MW1) and the second spin-echo $\pi/2$ pulse, as shown in the upper inset of Fig. \ref{dyn}. During this interval the population of $m_{s}=+1$ could relax, which can be readily measured using the NV-based magnetic resonance. While the population of $m_{s}=0$ is not accessible to direct measurement, it can be detected by driving the population to $m_{s}=+1$ state with a microwave pulse $\omega_{MW1}$. As a result, the relaxation duration $\tau_{rel}^{m_{s}=0}$ was inserted between the laser and microwave pulses, as shown in the lower inset of Fig. \ref{dyn}. By monitoring the amplitude of ESR peak as a function of $\tau_{rel}^{m_{s}=+1}$ and $\tau_{rel}^{m_{s}=0}$ measurements of two typical decay times of $T_{0}$-manifold, $T_{m_{s}=+1}$ and $T_{m_{s}=0}$, were obtained and shown in Fig. \ref{dyn}. Fitting to exponential curves, we obtained $T_{m_{s}=+1}=7\pm1$ $\mu$s and $T_{m_{s}=0}=23\pm5$ $\mu$s at $B_{0}=512$ G which are comparable with the traditional electron spin resonance (ESR) results within a factor of three \cite{6,7} (See the detailed description in Appendix \ref{comparison}). With the decay times of different molecular orientations, the kinetic parameters (lifetime and spin-lattice relaxation time) of the triplet sub-levels can be studied in a prescribed way \cite{6}.
\subsection{Polarization measurement}
The polarization $P$ is an important property for the remote quantum sensing. Experimentally, the sample was illuminated for 1.5 $\mu$s by a Gaussian beam with the waist radius $r_{0}$ = $35\pm5$ $\mu$m. Therefore, to quantitatively analyse the electron polarization of pentacene molecules, we integrate the sample volume modelled as a cylinder (V) with the thickness $h$ = $15\pm3$ $\mu$m and the radius $r_{0}$ = $35\pm5$ $\mu$m. In the weak signal limit, signal contrast is approximately equal to
\begin{equation}
\label{Eq:SC}
 \begin{aligned}
&SC \simeq 2\pi \gamma_e T_{m_S=+1}\overline{B}/2 \\
& = 2\pi \gamma_e T_{m_S=+1}/2*c^\prime\rho P\int_V\dfrac{3\cos^2(\theta)-1}{r^3} dV \\
& = 2\pi \gamma_e T_{m_S=+1}*\pi c^\prime\rho P(\dfrac{d+h}{\sqrt{r_0^2+(d+h)^2}}-\dfrac{d}{\sqrt{r_0^2+d^2}})
\end{aligned}
\end{equation}
where $T_{m_S=+1}$ = 7 $\mu$s is the lifetime of $m_S$ = +1, $\rho$ = 1.62$\times$10$^{-3}$ nm$^{-3}$ is the spin density of the sample, $c^\prime$ is a constant with the value 8.35 G$\cdot$nm$^{3}$ for pentacene spins (Note that the slope of $\omega^+_S$ is 2.53 MHz/G in the vicinity of 500 G, compared to gyromagnetic ratio 2.80 MHz/G for free electron spin giving $c= 9.28 G \cdot$nm$^{3}$ in the Eqn. \ref{2}), and $d$ is the NV depth below the diamond surface. $r$ and $\theta$ are the spherical coordinates in the frame taking the position of NV as the original point and the direction of magnetic field as the z axis.

\begin{figure}\centering
\includegraphics[width=1\columnwidth]{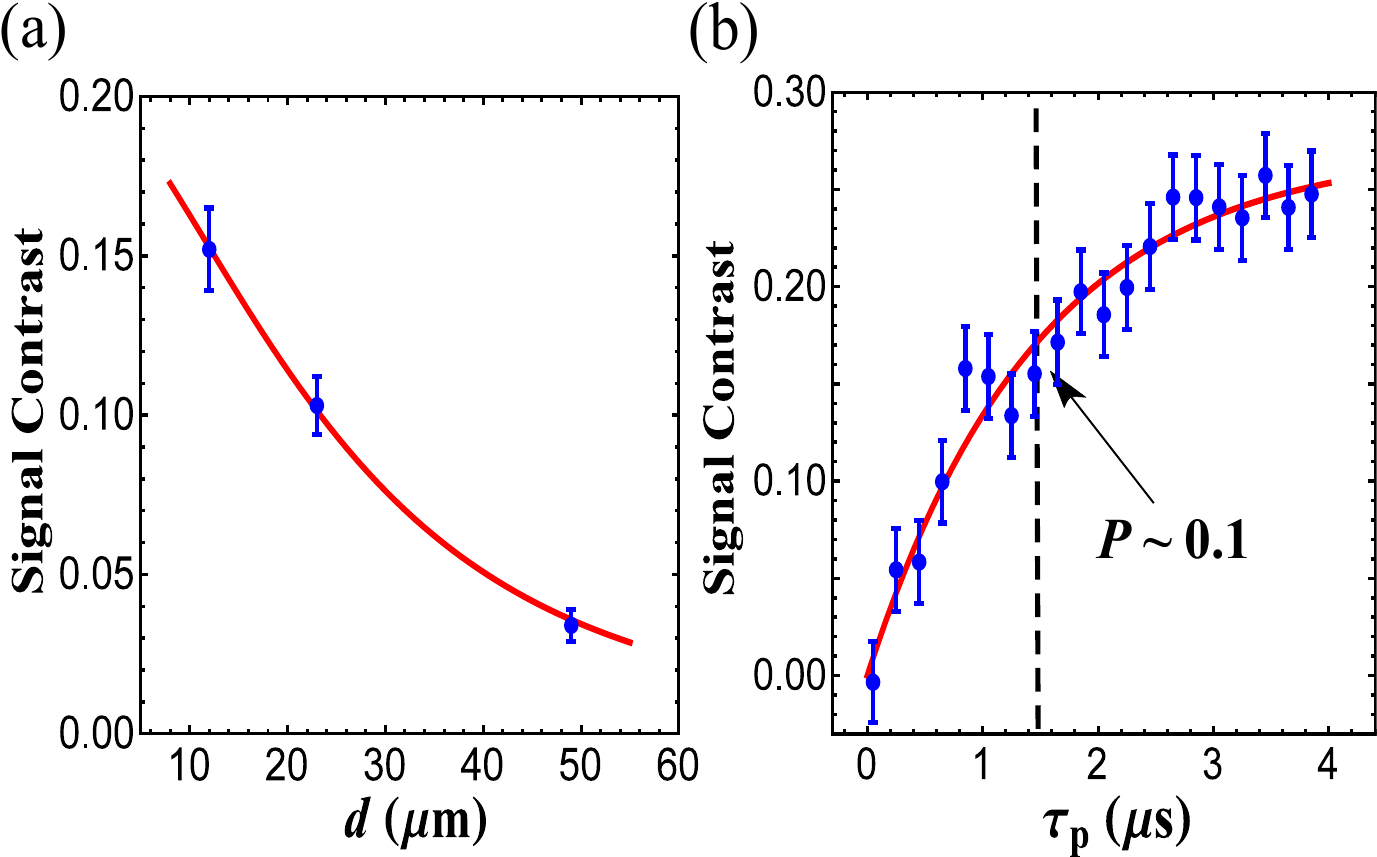}
\caption{(a) The peak intensity of OIPD as a function of NV depth. The signal recorded at three depths, $d=12, 23$ and $49$ $\mu$m. Other pulse sequence parameters: $\tau_{p}=$1.5 $\mu$s, $\tau=$30 $\mu$s. The solid line represents the curve-fitting where the cylindrical sample volume is decided by the radius $r=35\pm5$ $\mu$m and its thickness $h=15\pm3$ $\mu$m. Due to smaller signal detected by deeper NV, the pulse sequence was repeated more times to decrease uncertainty.
   (b) The magnetic resonance peak intensity as a function of laser pulse duration, $\tau_{p}$.
   }
    \label{lifetime}
\end{figure}

After integration of the sample volume, we curve-fit the data with Eqn. \ref{Eq:SC}, giving the polarization $P=0.10$, shown in Fig. \ref{lifetime} (a). Taking the uncertainties of $T_{m_{s}=+1}$, the radius and the thickness into consideration, we calculate the uncertainty of the polarization to be 0.01. The signal contrast decays with increasing the depth of NV owing to limited sample volume, unlike the case of infinite sample in Fig. \ref{range}(b). Larger polarzation can be achieved by increasing the 520nm-laser pulse duration. The PS signal is saturated with $\tau_{p}\geqslant$ 4 $\mu$s, indicating that the $T_{0}$ triplet manifold reaches equilibrium with the singlet state. A typical time $t_{l}=1.5\pm 0.3$ $\mu$s corresponding to the spin polarization of $P\sim 0.1$ is obtained by curve fitting, as shown in Fig. 6(b). The saturated polarization at 4 $\mu$s is $P\approx 0.19$ (the detailed comparison with conventional ESR experimental results is described in Appendix \ref{comparison}).

As we can see from Eq. \ref{4}, the sample thickness $h$ should be in the same order of magnitude or larger than the radius $r_{0}$. This is not a vital limitation of our method but just a technical issue. Generally, samples are prepared by cutting (for solid sample) or by designing a container, e.g. microfluidic devices, with a proper shape (for liquid sample). Moreover, gradient magnetic field can also be applied to selectively pick up the desired shape for detection.

\section{Conclusion}
These investigations were devised to develop techniques that enable the polarization signal to be probed with long-range NV sensing. Using the dipolar interactions of the NV center and a bulk magnetization, this approach has been achieved. With this long-range sensing protocol, we have studied the dynamical properties of pentacene molecule in a single crystal indicating broad applications in chemistry, biology and material science. Boltzmann-polarised spin magnetisation can be probed with NV-ensemble sensors as well \cite{9}. By manipulating NV sensors with different depths, we can vary the detection range of length scale from several hundred nanometers to ten microns to realize (sub-)cellular-sized magnetic resonance. With the improvement of polarization via hyperpolarization method, our approach can enable the applications of mesoscopic nuclear magnetic resonance spectroscopy and imaging at ambient conditions. 
\section*{Acknowledgments}
The authors thank Chang Gan Zeng for their help on the sample preparation, Haoli Zhang for useful discussion. This work is supported by the 973 Program (Grants No.~2013CB921800, No.~2016YFA0502400), the National Natural Science Foundation of China (Grants No.~11227901, No.~31470835, No.~91636217 and No.~11722544), the CAS (Grants No.~XDB01030400 and No.~QYZDY-SSW-SLH004, No.~YIPA2015370), the CEBioM, and the Fundamental Research Funds for the Central Universities (WK2340000064), the An Hui Natural Science Foundation (Grants No.~1708085MA22).

T. X, and F. S. contributed equally to this work.

\appendix

\section{Comparison with conventional ESR experiment}
\label{comparison}

\begin{figure}[H]
\includegraphics[width=1\columnwidth]{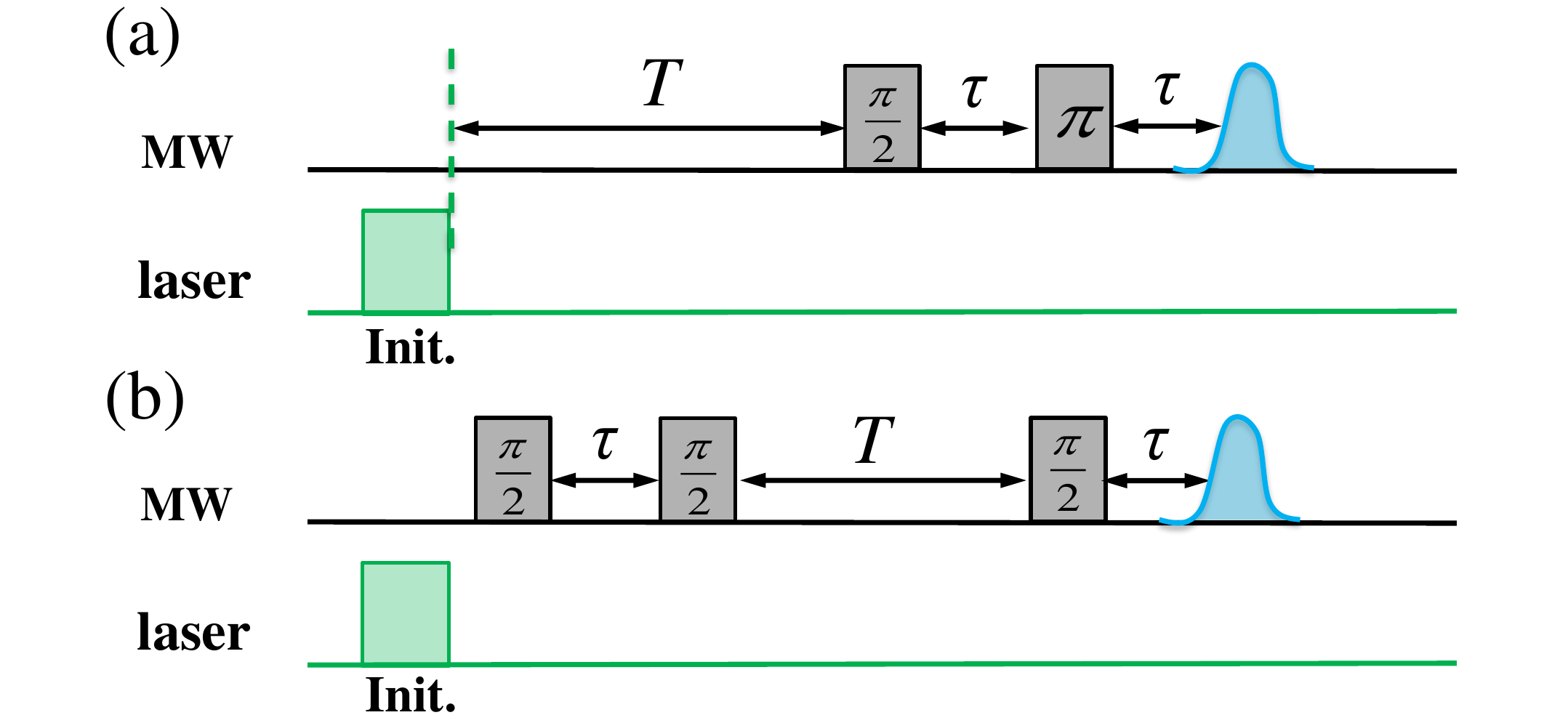}
\caption{Conventional ESR experiments measuring two decay times. (a) Two-pulse echo experiment pulse sequence. (b) Three-pulse ESEEM pulse sequence.}
\label{pulseq}
\end{figure}

Our measurements of two decay times $T_{m_{s}=0}$ and $T_{m_{s}=+1}$ reasonably agree with those of conventional ESR experiments. The decay time $T_{m_{s}=0}$ can be measured with two-pulse echo experiment the time between the $\pi$/2 and $\pi$ pulses is fixed, and the time delay between the laser pulse and the first microwave pulse T increases by steps, as shown in Fig. \ref{pulseq}(a). The time $\tau$ between the $\pi$/2 and $\pi$ microwave pulses was fixed at 1.25 $\mu$s. The result is shown in ref. \cite{7}, Fig. 2 (297 K) (about 30 $\mu$s), while in our work it is 23 $\mu$s.
The decay time $T_{m_{s}=+1}$ can be measured with three-pulse ($\pi$/2, $\pi$/2, $\pi$/2) ESEEM where the time delay between the second and third pulses T increases by steps, as shown in Fig. \ref{pulseq}(b). The first $\pi$/2 pulse occurred 1 $\mu$s after the laser pulse. The fixed delay time $\tau=0.36$ $\mu$s. The data is shown in ref. \cite{6}, Fig. 8(b) (about 15 $\mu$s), while in our work it is 7 $\mu$s. All the experiments performed in conventional ESR are above 3000 G while our experiments are performed at 500 G, and the pentacene density is also different from our sample. These differences may cause the difference between our results and conventional ESR experiments.\vspace{0.01 in}

In terms of polarization, the following factors may lead to the fact that the polarization ($\approx 0.19$) we measured is below the previous work ($\approx 0.6$). First, the laser pulse intensity in our experiment is at least two orders smaller than that in previous work, which renders polarization process so slow that it decays into singlet state. Second, as we mentioned in the text, strong transverse component $E$ of the ZFS mixes strongly spin sublevels $|0\rangle$ and $|\pm1\rangle$ at the magnetic field of $\sim 500$ G. Therefore, the population distribution is different from that in the previous work.

\end{document}